\documentclass[fleqn,11pt,twoside]{article}
\usepackage{amsthm}
\usepackage{amsmath}
\usepackage{graphicx}
\usepackage{latexsym}
\makeatletter

\newcommand{\Name}[1]{\begin{flushleft}
                       \LARGE \bf #1
                       \end{flushleft}\vspace{-3mm}}

\newcommand{\Author}[1]{\begin{flushleft}
                       \it #1 \end{flushleft}}

\newcommand{\Address}[1]{\begin{flushleft}
                       \it #1 \end{flushleft}}

\newcommand{\FirstPageHead}[5]{
\begin{flushleft}
\raisebox{8mm}[0pt][0pt]
{\footnotesize \sf
\parbox{150mm}{ \qquad
 #1 #2 #3 
#4\hfill {\sc #5}}}\vspace{-13mm}
\end{flushleft}}

\newcommand{\evenhead}{Author \ name}
\newcommand{\oddhead}{Article \ name}

\renewcommand{\@evenhead}{
\hspace*{-3pt}\raisebox{-15pt}[\headheight][0pt]{\vbox{\hbox to \textwidth
{\thepage \hfil \evenhead}\vskip4pt \hrule}}}
\renewcommand{\@oddhead}{
\hspace*{-3pt}\raisebox{-15pt}[\headheight][0pt]{\vbox{\hbox to \textwidth
{\oddhead \hfil \thepage}\vskip4pt\hrule}}}
\renewcommand{\@evenfoot}{}
\renewcommand{\@oddfoot}{}

\long\def\@makecaption#1#2{%
  \vskip\abovecaptionskip
  \sbox\@tempboxa{\small \textbf{#1.}\ \ #2}%
  \ifdim \wd\@tempboxa >\hsize
    {\small \textbf{#1.}\ \ #2}\par
  \else
    \global \@minipagefalse
    \hb@xt@\hsize{\hfil\box\@tempboxa\hfil}%
  \fi
  \vskip\belowcaptionskip}

\newcommand{\JNMPnumberwithin}[3][\arabic]{%
  \@ifundefined{c@#2}{\@nocounterr{#2}}{%
    \@ifundefined{c@#3}{\@nocnterr{#3}}{%
      \@addtoreset{#2}{#3}%
      \@xp\xdef\csname the#2\endcsname{%
        \@xp\@nx\csname the#3\endcsname .\@nx#1{#2}}}}%
}

\newcommand{\resetfootnoterule} {
  \renewcommand\footnoterule{%
  \kern-3\p@
  \hrule\@width.4\columnwidth
  \kern2.6\p@}
}

\renewcommand{\footnoterule}{}

\newcommand{\be}{\begin{equation}}
\newcommand{\ee}{\end{equation}}
\newcommand{\ba}{\hspace*{-5pt}\begin{array}}
\newcommand{\ea}{\end{array}}
\newcommand{\p}{\partial}

\makeatother

\numberwithin{equation}{section}
\theoremstyle{definition}

\renewcommand{\ba}{\begin{array}}
\renewcommand{\ea}{\end{array}}
\newcommand{\beg}{\begin{eqnarray}}
\newcommand{\eeq}{\end{eqnarray}}
\newcommand{\bg}{\begin{eqnarray*}}

\newcommand{\ed}{\end{eqnarray*}}
\newcommand{\n}{\newline\hfill}

\renewcommand{\p}{\partial} 
 
\newcommand{\notlhd}{\lhd\kern-.8em{/}\ } 
\newcommand{\notexist}{\ \exists\kern-.5em{\raise.1em\hbox{/}}\ }

\newcommand{\pde}[2]{\frac{\p #1}{\p #2}} 
 
\newcommand{\pdd}[2]{\frac{\p^2 #1}{\p #2^2}} 
\newcommand{\inp}{{\mbox{\vbox{\hrule width0ex\hbox{\vrule
 height0ex\kern3.8pt
\vbox{\kern2.5pt}\kern3.8pt \vrule height1.6ex}
\hrule width1.6ex}}}}

\begin{document}

\renewcommand{\evenhead}{N Euler and M Euler}
\renewcommand{\oddhead}{Nonlocal symmetries, conservation laws and 
transformations}


\thispagestyle{empty}

\begin{flushleft}
\footnotesize \sf
\end{flushleft}

\FirstPageHead{\ }{\ }{\ }
{ }{{
{ }}}

\Name{On nonlocal symmetries, nonlocal conservation laws
and nonlocal transformations of evolution equations: Two
linearisable hierarchies}



\Author{\bf Norbert Euler and Marianna Euler}



\Address{
Department of Mathematics,  
Lule\aa\ University of Technology \\
SE-971 87 Lule\aa, Sweden\\
Norbert.Euler@sm.luth.se; Marianna.Euler@sm.luth.se
}

\vspace{1cm}

\noindent
{\bf Abstract}:
We discuss nonlocal symmetries and nonlocal conservation laws that
follow from the systematic potentialisation of evolution equations.
Those are the Lie point symmetries
of the auxiliary systems, also known as potential symmetries.
We define higher-degree
potential symmetries which then lead to nonlocal conservation laws and
nonlocal transformations for the equations.
We demonstrate our approach and derive second degree potential
symmetries for the Burgers' hierarchy and the
Calogero-Degasperis-Ibragimov-Shabat hierarchy.


\section{Introduction}
The concept of nonlocal symmetries of partial differential equations
and its relations to local Lie point symmetries of its associated
auxiliary systems was introduced by Bluman, Kumei and Reid \cite{BKR}
and are known as potential symmetries. 
We point out that more general type of potential symmetries
were introduced earlier by Krasil'shchik and Vinogradov (see \cite{KV}
and \cite{VK-1984}).

In the present paper our starting point is based on  
potential symmetries as introduced in \cite{BKR} and
\cite{Bluman_Kumei}.
We define higher-degree nonlocal
symmetries for evolution equations by introducing further
auxiliary system by higher-degree potentialisations.
This leads to nonlocal conservation laws for the given evolution
equations and to
nonlocal transformations between the evolution equations and its
potentialised equations.
We demonstrate our approach by considering the well-known
Burgers' hierarchy and the so-called Calogero-Degasperis-Ibragimov-Shabat
hierarchy (\cite{Calogero_Degasperis}, \cite{Ibragimov_Shabat},
(\cite{Sokolov84}, (\cite{Calogero87}). Both of these hierarchies
are known to be linearisable (see e.g. \cite{Calogero87},
\cite{MNP} and \cite{Petersson}. See also \cite{S_S} for a
discussion on nonlocal symmetries of the
Calogero-Degasperis-Ibragimov-Shabat equation).
We show that the linearisations of the two hierarchies follow directly
from their second potentialisations. An interesting and unexpected
result of our investigation is that second-degree potential
symmetries (as defined by Definition 2.1) exist only for the first
members of both the Burgers' and the Calogero-Degasperis-Ibragimov-Shabat
hierarchies. 

{\it On the notation:} Throughout this paper $D_a[p]$ denotes the total
derivative-operator of the dependent variable $p(a,b)$ with respect to the
independent variable $a$, where subscripts of
$p$ denote partial derivatives:
\begin{gather}
D_a[p]:=\pde{\ }{a}+p_a\,\pde{\ }{p}+p_{aa}\,\pde{\ }{p_a}+
p_{ab}\,\pde{\ }{p_b}+
p_{3a}\,\pde{\ }{p_{aa}}+\cdots\ .
\end{gather}
The formal inverse-operator of $D_a[p]$ is denoted by $D_a^{-1}$,
such that
\begin{gather}
D_a^{-1}\circ D_a[p]\,\varphi=D_a[p]\circ D_a^{-1}\,\varphi=\varphi.
\end{gather}
Moreover
\begin{gather}
D^n_a[p]\,\varphi=D^{n-1}_a[p]\circ D_a[p]\,\varphi,\qquad n\in {\cal N}.
\end{gather}
If the dependence of the operator $D_a[p]$ on $p$ is obvious,
we write just $D_a$ instead of $D_a[p]$.

\section{Preliminaries and higher-degree potential symmetries}

Consider an $n$th-order evolution equation of the general form
\begin{gather}
\label{evolution}
u_t=F(x,u,u_x,u_{xx},u_{3x},\ldots, u_{nx}).
\end{gather}
Assume that (\ref{evolution}) is a symmetry-integrable evolution
equation \cite{Fokas}, i.e. (\ref{evolution}) admits a
hereditary recursion operator $R[u]$ such that
\begin{gather}
\left[L_F[u],\ R[u]\right]=D_t[u]\,R[u],
\end{gather}
where $L_F[u]$ is the linear operator
\begin{gather}
L_F[u]:=\pde{F}{u}+\pde{F}{u_x}D_x+\pde{F}{u_{xx}}D_x^2+\cdots\ .
\end{gather}
Assume further that the hierarchy of symmetry-integrable evolution
equations can
be presented in the form
\begin{gather}
\label{Int_hier}
u_t=R^n[u]\,u_x,\qquad n\in {\cal N}, 
\end{gather}
such that (\ref{evolution}) corresponds to the first member of the
hierarchy (\ref{Int_hier}) with $n=1$.
The conserved current, $\Phi^t$,
for (\ref{evolution}) must satisfy the relation
(\cite{Fokas_Fuchssteiner}, \cite{Anco})
\begin{gather}
\label{Int_Phi^t}
\Lambda=\hat E[u]\,\Phi^t,
\end{gather}
where $\Lambda$ denotes an integrating factor for (\ref{evolution}),
i.e.
\begin{gather}
\label{Int_lamb}
\hat E[u]\left(\Lambda u_t-\Lambda F(x,u,u_x,u_{xx},\ldots
  u_{nx})\right)=0.
\end{gather}
Here $\hat E[u]$ is the Euler operator
\begin{gather}
\hat E[u]:=\pde{\ }{u}-D_t\circ \pde{\ }{u_t}-D_x\circ \pde{\ }{u_x}+
D_x^2\circ \pde{\ }{u_{xx}}-D_x^3\circ \pde{\ }{u_{3x}}+\cdots\ .
\end{gather}

\strut\hfill

For the flux, $\Phi^x$, we state

\strut\hfill

\noindent
{\bf Proposition 2.1:} {\it
Let $\Lambda$ be an integrating factor for the evolution equations 
(\ref{evolution}) and assume that the corresponding conserved current, 
$\Phi^t$, admits the dependence
\begin{gather}
\label{assume_dep}
\Phi^t=\Phi^t(x,u,u_x,u_{xx},u_{3x}).
\end{gather}
Then the flux, $\Phi^x$, for (\ref{evolution}) is given by
\begin{gather}
\Phi^x=-D_x^{-1}\left(\Lambda\,F\right)-\pde{\Phi^t}{u_x}F
-\pde{\Phi^t}{u_{xx}}D_xF
-\pde{\Phi^t}{u_{3x}}D_x^2F\nonumber\\[0.3cm]
\label{Int_Phi^x}
\qquad+FD_x\left(\pde{\Phi^t}{u_{xx}}\right)
-FD_x^2\left(\pde{\Phi^t}{u_{3x}}\right)
+(D_xF)D_x\left(\pde{\Phi^t}{u_{3x}}\right).
\end{gather}
The hierarchy (\ref{Int_hier}) admits the same
integrating factor,  
$\Lambda$, as the first member of the hierarchy (for $n=1$) and hence the
same corresponding current, $\Phi^t$. The flux, $\Phi^x$, for the 
hierarchy, (\ref{Int_hier}), for all $n\in {\cal N}$ then takes the fom
\begin{gather}
\Phi^x(x,u,u_x,\ldots;n)=-D_x^{-1}\left(\Lambda\,R^n[u]\,F\right)-\pde{\Phi^t}{u_x}R^n[u]\,F
-\pde{\Phi^t}{u_{xx}}D_x\left(R^n[u]\,F\right)\nonumber\\[0.3cm]
\qquad -\pde{\Phi^t}{u_{3x}}D_x^2\left(R^n[u]\,F\right)
+\left(R^n[u]\,F\right)D_x\left(\pde{\Phi^t}{u_{xx}}\right)
-\left(R^n[u]\,F\right)D_x^2\left(\pde{\Phi^t}{u_{3x}}\right)
\nonumber\\[0.3cm]
\qquad+D_x\left(R^n[u]\,F\right)D_x\left(\pde{\Phi^t}{u_{3x}}\right),
\end{gather}
where we assume the dependence of $\Phi^t$ as stated in
(\ref{assume_dep}).
}

\strut\hfill

\noindent
{\bf Remark}:
The proof of Proposition 2.1 is straightforward, namely by integrating the 
conservation law
\begin{gather}
\left.\left(D_t\Phi^t+D_x\Phi^x\right)\right|_{u_t=R^{n}[u]u_x}=0
\end{gather}
of the hierarchy (\ref{Int_hier}) with respect to $x$, that is
\begin{gather*}
\left.\Phi^x=-D_x^{-1}\left(D_t\Phi^t\right)\right|_{u_t=R^{n}[u]u_x}.
\end{gather*}

Assume now that the evolution equation (\ref{evolution}) admits 
a conserved current, $\Phi_1^t$, and flux, $\Phi_1^x$. Following
\cite{BKR} a {\it first potential
variable} $v$ is then defined by the auxiliary system:
\begin{subequations}
\begin{gather}
\label{Int_vx}
v_x=\Phi_1^t(x,u,u_x,\ldots)\\[0.3cm]
\label{Int_vt}
v_t=-\Phi_1^x(x,u,u_x,\ldots).
\end{gather}
\end{subequations}
We name system (\ref{Int_vx})--(\ref{Int_vt}) the
{\it first auxiliary system} of (\ref{evolution}).
Assume further that (\ref{Int_vt})
can be expressed in terms of the first potential variable $v$, i.e.
(\ref{Int_vt}) becomes by (\ref{Int_vx})
the {\it first potential equation} of the general form
\begin{gather}
\label{Int_pot_eq_1}
v_t=G(x,v_x,v_{xx},\ldots,v_{nx})
\end{gather}
which may again admit a conserved current, $\Phi_2^t$, and flux, $\Phi_2^x$.
A further potential $w$ is then introduced for (\ref{Int_pot_eq_1}),
and named the {\it second potential} for (\ref{evolution}), by
the {\it second auxiliary system}
\begin{subequations}
\begin{gather}
\label{Int_wx}
w_x=\Phi_2^t(x,v,v_x,\ldots)\\[0.3cm]
\label{Int_wt}
w_t=-\Phi_2^x(x,v,v_x,\ldots).
\end{gather}
\end{subequations}
The corresponding potential equation for (\ref{Int_pot_eq_1})
is then obtained from (\ref{Int_wx}) and  (\ref{Int_wt}),
which we assume to have the general form 
\begin{gather}
\label{Int_pot_eq_2}
w_t=H(x,w_x,w_{xx},\ldots,w_{nx}).
\end{gather}
We name (\ref{Int_pot_eq_2}) the {\it second potential equation} for
(\ref{evolution}).

\strut\hfill

We now introduce the following

\strut\hfill

\noindent
{\bf Definition 2.1:} {\it The Lie point symmetry generators
\begin{gather}
Z=\xi_1(x,t,u,v)\pde{\ }{x}+\xi_2(x,t,u,v)\pde{\ }{t}
+\eta_1(x,t,u,v)\pde{\ }{u}+\eta_2(x,t,u,v)\pde{\ }{v}
\end{gather}
of the first auxiliary system (\ref{Int_vx})--(\ref{Int_vt})
for (\ref{evolution}),
i.e.
\begin{gather*}
v_x=\Phi_1^t(x,u,u_x,\ldots)\\[0.3cm]
v_t=-\Phi_1^x(x,u,u_x,\ldots),
\end{gather*}
are defined as the first-degree potential symmetries of
(\ref{evolution})
if the infinitesimals $\xi_1,\ \xi_2$ and $\eta_1$ depend essentially
on the first potential variable $v$, that is
\begin{gather}
\left(\pde{\xi_1}{v}\right)^2+\left(\pde{\xi_2}{v}\right)^2
+\left(\pde{\eta_1}{v}\right)^2
\neq 0.
\end{gather}
The second-degree potential symmetries of (\ref{evolution})
are defined by the Lie point symmetry generators of the combined
first- and second-auxiliary systems (\ref{Int_vx})--(\ref{Int_vt})
and (\ref{Int_wx})--(\ref{Int_wt}), that is the Lie point symmetry
generators of the form
\begin{gather}
Z=\xi_1(x,t,u,v,w)\pde{\ }{x}+\xi_2(x,t,u,v,w)\pde{\ }{t}
+\eta_1(x,t,u,v,w)\pde{\ }{u}\nonumber\\[0.3cm]
\qquad +\eta_2(x,t,u,v,w)\pde{\ }{v}
+\eta_3(x,t,u,v,w)\pde{\ }{w}
\end{gather}
for the system
\begin{gather*}
v_x=\Phi_1^t(x,u,u_x,\ldots)\\[0.3cm]
v_t=-\Phi_1^x(x,u,u_x,\ldots)\\[0.3cm]
w_x=\Phi_2^t(x,v,v_x,\ldots)\\[0.3cm]
w_t=-\Phi_2^x(x,v,v_x,\ldots),
\end{gather*}
where the infinitesimals $\xi_1,\ \xi_2$, $\eta_1$ and $\eta_2$
depend essentially on the second potential variable $w$,
that is
\begin{gather}
\left(\pde{\xi_1}{w}\right)^2+\left(\pde{\xi_2}{w}\right)^2
+\left(\pde{\eta_1}{w}\right)^2+\left(\pde{\eta_2}{w}\right)^2
\neq 0
\end{gather}
}

It should be clear that Definition 2.1 can easily be extended to
$m$th-degree potential symmetries.

\section{The Burgers' hierarchy}
Consider the Burgers' equation in the form
\begin{gather}
\label{Burgers}
u_t=u_{xx}+2uu_x.
\end{gather}
It is well known that (\ref{Burgers}) admits only one local integrating
factor and one local conservation law (see e.g. \cite{Olver}), where
\begin{gather}
\Lambda=1,\qquad \Phi_1^t=u,\qquad \Phi_1^x=-\left(u_x+u^2\right).
\end{gather}
Equation (\ref{Burgers}) admits the recursion operator \cite{Oevel}
\begin{gather}
\label{Burg_R}
R[u]=D_x+u+u_xD_x^{-1}\circ 1
\end{gather}
and the Burgers' hierarchy then takes the form
\begin{gather}
\label{Burg_hier}
u_t=R^n[u]\, u_x,\qquad n=1,2,\ldots \ .
\end{gather}
We remark that a general class of linearisable second-order evolution
equations and its recursion operators, for which the Burgers'
hierarchy is a special case, was reported in \cite{NM} and \cite{MNP}.

\subsection{Nonlocal conservation laws and linearisation}

We prove the following

\strut\hfill

\noindent
{\bf Proposition 3.1:} {\it The Burgers' hierarchy (\ref{Burg_hier}),
\begin{gather*}
u_t=R^n[u]\,u_x,\qquad n=1,2,\ldots\ , 
\end{gather*}
with $R$ given by (\ref{Burg_R}) admits the first
potentialisation of the form
\begin{gather}
\label{Burg_pot_1}
v_t=P^n[v_x]\,v_x,\qquad n=1,2,\ldots\ ,
\end{gather}
where
\begin{gather}
P[v_x]=D_x[v_x]+v_x,
\end{gather}
and the second potentialisation
\begin{gather}
\label{Burg_lin}
w_t=w_{(n+1)x},\qquad n=1,2,\ldots\ ,
\end{gather}
where
\begin{subequations}
\begin{gather}
\label{Burg_aux_1a}
v_x=u\\[0.3cm]
\label{Burg_aux_1b}
v_t=P^n[u]\,u,\qquad n=1,2,\ldots\ ,\\[0.3cm]
\label{Burg_aux_2a}
w_x=e^v\\[0.3cm]
\label{Burg_aux_2b}
w_t=D^n_x[v]\,e^v,\qquad n=1,2,\ldots\ ,
\end{gather}
\end{subequations}
and
\begin{gather}
\label{P}
P[u]=D_x+u.
\end{gather}
The corresponding nonlocal conserved current, $\Phi^t$, and
flux, $\Phi^x$, for 
hierarchy (\ref{Burg_hier}) are
\begin{subequations}
\begin{gather}
\label{Burg-Phi-t}
\Phi^t=e^{\int u\,dx}\\[0.3cm]
\label{Burg-Phi-x}
\Phi^x=-D_x^n[u]\left(e^{\int u\,dx}\right),\qquad
n=1,2,\ldots\ 
\end{gather}
\end{subequations}
and the linearising transformation that transforms (\ref{Burg_hier}) 
in (\ref{Burg_lin}) is
\begin{gather}
\label{Burg_trans}
w_x=e^{\int u \,dx}.
\end{gather}

}

\noindent
{\bf Proof:}
By Proposition 2.1 the hierarchy (\ref{Burg_hier})
admits the following integrating factor, $\Lambda$,
conserved current, $\Phi_1^t$, and flux, $\Phi_1^x$:
\begin{subequations}
\begin{gather}
\label{Lamb_1}
\Lambda=1\\[0.3cm]
\label{Burg-Phi^t_1}
\Phi_1^t(u)=u\\[0.3cm]
\label{Burg-Phi^x_1}
\Phi^x_{1,n}=-D_x^{-1}\left(R^n[u]u_x\right),\qquad
n=1,2,\ldots\ ,
\end{gather}
\end{subequations}
where $R$ is the recursion operator, (\ref{Burg_R}).
It is easy to verify that
\begin{gather}
\label{Burg-R-P}
D_x^{-1}\left(R^n[u]\,u_x\right)=P^n[u]\,u,
\qquad n=1,2,\ldots\ ,
\end{gather}
where $P$ is defined by (\ref{P}).
The first auxiliary system 
for the Burgers' hierarchy (\ref{Burg_hier}) is then defined 
in terms of a potential variable $v$ in the form
(\ref{Burg_aux_1a})--(\ref{Burg_aux_1b}), i.e.
\begin{subequations}
\begin{gather*}
v_x=u\\[0.3cm]
v_t=P^n[u]\,u,\qquad n=1,2,\ldots\ ,
\end{gather*}
\end{subequations}
and the first potential hierarchy of the Burgers' hierarchy,
(\ref{Burg_hier}),
becomes (\ref{Burg_pot_1}), i.e.
\begin{gather*}
v_t=P^n[v_x]\,v_x,\qquad n=1,2,\ldots\ .
\end{gather*}
The first potential hierarchy, (\ref{Burg_pot_1}),
admits the following integrating factor, $\Lambda$, conserved current,
$\Phi^t$, and flux, $\Phi^x$:
\begin{subequations}
\begin{gather}
\Lambda=e^v\\[0.3cm]
\Phi_2^t(v)=e^v\\[0.3cm]
\Phi_{2,n}^x=-D_x^{-1}\left(e^vP^n[v_x]\,v_x\right),
\qquad n=1,2,\ldots\ .
\end{gather}
\end{subequations} 
By the relation
\begin{gather}
D_x^{-1}\left(e^vP^n[v_x]\,v_x\right)=D^n_x[v]\,e^v,
\qquad  n=1,2,\ldots\ ,
\end{gather}
the second auxiliary system of the Burgers' hierarchy (\ref{Burg_hier})
is then defined in the form (\ref{Burg_aux_2a})--(\ref{Burg_aux_2b}),
i.e.
\begin{subequations}
\begin{gather*}
w_x=e^v\\[0.3cm]
w_t=D^n_x[v]\,e^v,\qquad n=1,2,\ldots\ ,
\end{gather*}
\end{subequations}
so that the second potential hierarchy becomes
\begin{gather*}
w_t=D_x^n[w]\,w_{x}\equiv w_{(n+1)x},\qquad n=1,2,\ldots\ .
\end{gather*}
The nonlocal transformation (\ref{Burg_trans})
follows directly from (\ref{Burg_aux_1a}) and
(\ref{Burg_aux_2a}), namely the
well known Cole-Hopf transformation (see e.g. \cite{Olver}).
The nonlocal conserved current (\ref{Burg-Phi-t}) and
flux (\ref{Burg-Phi-x}) follows directly by expressing 
(\ref{Burg_aux_2a})--(\ref{Burg_aux_2b}) in terms of the original
variable $u$. 
\hfill{$\Box$}

\subsection{Potential symmetries of the Burgers' hierarchy}

We now turn our attention to the symmetry properties of the
auxiliary systems (\ref{Burg_aux_1a})--(\ref{Burg_aux_1b})
and the combined auxiliary system (\ref{Burg_aux_1a})--(\ref{Burg_aux_2b}).
Firstly we discuss in detail the cases $n=1$ and $n=2$.

\strut\hfill

\noindent
{\bf Case} $n=1$: The first auxiliary system 
(\ref{Burg_aux_1a})--(\ref{Burg_aux_1b}) of hierarchy
(\ref{Burg_hier})
with $n=1$ is
\begin{subequations}
\begin{gather}
\label{B-n0-A1a}
v_x=u\\
\label{B-n0-A1b}
v_t=u_x+u^2
\end{gather}
\end{subequations}
and the first  potential equation has the form
\begin{gather}
v_t=v_{xx}+v_x^2.
\end{gather}
By Definition 2.1 the first-degree potential symmetries of
(\ref{Burgers}) are the
Lie point symmetries of (\ref{B-n0-A1a})--(\ref{B-n0-A1b}). We obtain
\begin{subequations}
\begin{gather}
\label{Burg_n1_sym_A}
Z_1=\pde{\ }{t},\qquad Z_2=\pde{\ }{x}, \qquad Z_3=\pde{\ }{v}\\[0.3cm]
Z_4=x\pde{\ }{x}+2t\pde{\ }{t}-u\pde{\ }{u},\qquad
Z_5=2t\pde{\ }{x}-\pde{\ }{u}-x\pde{\ }{v}\\[0.3cm]
Z_6=4xt\pde{\ }{x}+4t^2\pde{\ }{t}-2(x+2tu)\pde{\ }{u}
-(2t+x^2)\pde{\ }{v}\\[0.3cm]
\label{Burg_n1_sym_B}
Z_\infty=e^{-v}\left(\pde{f}{x}-uf(x,t)\right)\pde{\ }{u}
+f(x,t)e^{-v}\pde{\ }{v},\qquad \mbox{where}\ \ f_t-f_{xx}=0.
\end{gather}
\end{subequations}
The first-degree potential symmetries
(\ref{Burg_n1_sym_A})--(\ref{Burg_n1_sym_B})
were firstly obtained by Vino'gradov and Krasil'shchik \cite{VK-1984}.  
The second auxiliary system (\ref{Burg_aux_2a})--(\ref{Burg_aux_2b})
for hierarchy (\ref{Burg_hier}) with $n=1$ is
\begin{subequations}
\begin{gather}
\label{B-n0-A2a}
w_x=e^v\\
\label{B-n0-A2b}
w_t=v_xe^v
\end{gather}
\end{subequations}
and the second potential equation for hierarchy (\ref{Burg_hier}) with $n=1$ 
has the form
\begin{gather}
w_t=w_{xx}.
\end{gather}
Following Definition 2.1 the second-degree potential symmetries of the
Burgers hierarchy
(\ref{Burg_hier}) for $n=1$ are the Lie point symmetries
of the combined auxiliary systems (\ref{B-n0-A1a})--(\ref{B-n0-A1b})
and (\ref{B-n0-A2a} )--(\ref{B-n0-A2b}), i.e. the Lie point symmetries
of the system
\begin{subequations}
\begin{gather}
\label{NB-n0-A1a}
v_x=u\\
\label{NB-n0-A1b}
v_t=u_x+u^2\\
\label{NB-n0-A2a}
w_x=e^v\\
\label{NB-n0-A2b}
w_t=v_xe^v.
\end{gather}
\end{subequations}
We obtain the following second-degree potential symmetries of 
(\ref{Burg_hier}) for $n=1$:
\begin{subequations}
\begin{gather}
Z_1=\pde{\ }{t},\qquad Z_2=\pde{\ }{x}\\[0.3cm]
Z_3=x\pde{\ }{x}+2t\pde{\ }{t}-u\pde{\ }{u}+w\pde{\ }{w},\qquad
Z_4=w\pde{\ }{w}+\pde{\ }{v}\\[0.3cm]
Z_5=2t\pde{\ }{x}
-\left(2-uwe^{-v}\right)\pde{\ }{u}
-\left(x+we^{-v}\right)\pde{\ }{v}
-xw\pde{\ }{w}
\end{gather}
\end{subequations}

\begin{subequations}
\begin{gather}
Z_6=2xt\pde{\ }{x}
+2t^2\pde{\ }{t}
-\left(2x+2tu+we^{-v}-xuwe^{-v}\right)\pde{\ }{u}\\[0.3cm]
\qquad
-\left(3t+\frac{1}{2}x^2+xwe^{-v}\right)\pde{\ }{v}
-\left(tw+\frac{1}{2}x^2w\right)\pde{\ }{w}\\[0.3cm]
Z_\infty=
e^{-v}\left(u\pde{f}{x}-\pdd{f}{x}\right)\pde{\ }{u}
-e^{-v}\pde{f}{x}\pde{\ }{v}
-f(x,t)\pde{\ }{w},\\[0.3cm]
\mbox{where}\ \  f_t-f_{xx}=0.\nonumber
\end{gather}
\end{subequations}

\strut\hfill

\noindent
{\bf Case $n=2$:} The first auxiliary system 
(\ref{Burg_aux_1a})--(\ref{Burg_aux_1b}) of hierarchy (\ref{Burg_hier})
with $n=2$ is
\begin{subequations}
\begin{gather}
\label{B-n2-A1a}
v_x=u\\
\label{B-n2-A1b}
v_t=u_{xx}+3uu_x+u^3
\end{gather}
\end{subequations}
and the first potential equation has the form
\begin{gather}
\label{Burgers_3}
v_t=v_{3x}+3v_xv_{xx}+v_x^3.
\end{gather}
The first-degree potential symmetries of (\ref{Burgers}) are
then
\begin{subequations}
\begin{gather}
Z_1=\pde{\ }{t},\qquad Z_2=\pde{\ }{x}, \qquad Z_3=\pde{\ }{v}\\[0.3cm]
Z_4=x\pde{\ }{x}+3t\pde{\ }{t}-u\pde{\ }{u}\\[0.3cm]
Z_\infty=e^{-v}\left(\pde{f}{x}-uf(x,t)\right)\pde{\ }{u}
+f(x,t)e^{-v}\pde{\ }{v},\\[0.3cm]
\mbox{where}\ \
f_t-f_{3x}=0.\nonumber
\end{gather}
\end{subequations}
The second auxiliary system (\ref{Burg_aux_2a})--(\ref{Burg_aux_2b})
for hierarchy (\ref{Burg_hier}) with $n=2$ is
\begin{subequations}
\begin{gather}
\label{B-n2-A2a}
w_x=e^v\\
\label{B-n2-A2b}
w_t=v_{xx}e^v+v_x^2e^v
\end{gather}
\end{subequations}
and the second potential equation for hierarchy (\ref{Burg_hier}) with
$n=2$ has the form
\begin{gather}
w_t=w_{3x}.
\end{gather}
The second-degree potential symmetries of the Burgers' hierarchy
(\ref{Burg_hier}) for $n=2$ would follow from the Lie point symmetries
of the combined auxiliary systems (\ref{B-n0-A1a})--(\ref{B-n0-A1b})
and (\ref{B-n0-A2a} )--(\ref{B-n0-A2b}), i.e. the Lie point symmetries
of the system
\begin{subequations}
\begin{gather}
\label{NB-n2-A1a}
v_x=u\\
\label{NB-n2-A1b}
v_t=u_{xx}+3uu_x+u^3\\
\label{NB-n2-A2a}
w_x=e^v\\
\label{NB-n2-A2b}
w_t=v_{xx}e^v+v_x^2e^v.
\end{gather}
\end{subequations}
We obtain the following Lie point symmetries of 
system (\ref{NB-n2-A1a})--(\ref{NB-n2-A2b}):
\begin{subequations}
\begin{gather}
Z_1=\pde{\ }{t},\qquad Z_2=\pde{\ }{x}\\[0.3cm]
Z_3=x\pde{\ }{x}+3t\pde{\ }{t}-u\pde{\ }{u}+w\pde{\ }{w},\qquad
Z_4=w\pde{\ }{w}+\pde{\ }{v}\\[0.3cm]
Z_\infty=
e^{-v}\left(u\pde{f}{x}-\pdd{f}{x}\right)\pde{\ }{u}
-e^{-v}\pde{f}{x}\pde{\ }{v}
-f(x,t)\pde{\ }{w},\\[0.3cm]
\mbox{where}\ \
f_t-f_{3x}=0.\nonumber
\end{gather}
\end{subequations}

\strut\hfill

\noindent
It is clear that the above Lie point symmetry generators are not
potential symmetries of second degree for the third-order Burgers'
equation (\ref{Burgers_3}). The same happens for the case $n=3$,
i.e., second-degree potential symmetries for the Burgers' hierarchy
appear only for the case $n=1$, namely the Burgers' equation
(\ref{Burgers}).

\strut\hfill

From the above patterns in the symmetry generators we allow ourselves
the following

\strut\hfill

\noindent
{\bf Supposition 3.1:} {\it
There exist no second-degree potential symmetries for the
Burgers' hierarchy
(\ref{Burg_hier}) for $n>1$ and
the maximum set of first-degree potential symmetries for 
the hierarchy (\ref{Burg_hier})
for all natural numbers $n>1$,
is given by the following Lie symmetry generators:
\begin{subequations}
\begin{gather}
Z_1=\pde{\ }{t},\qquad
Z_2=\pde{\ }{x},\qquad
Z_3=\pde{\ }{v}\\[0.3cm]
Z_4=x\pde{\ }{x}+(n+1)t\pde{\ }{t}-u\pde{\ }{u}\\[0.3cm]
Z_\infty=
e^{-v}\left(\pde{f}{x}-uf(x,t)\right)\pde{\ }{u}
+f(x,t)e^{-v}\pde{\ }{v},\\[0.3cm]
\mbox{where}\ \ f_t-f_{(n+1)x}=0.\nonumber
\end{gather}
\end{subequations}
}

\subsection{Reciprocal-B\"acklund transformations of the Burgers' hierarchy}

With the general expressions of the conserved current,
(\ref{Burg-Phi^t_1}),
and flux, (\ref{Burg-Phi^x_1}), together with the relation
(\ref{Burg-R-P}) for the the Burgers' hierarchy (\ref{Burg_hier}), we
use the opportunity to transform the hierarchy by a
reciprocal-B\"acklund transformation (see e.g. \cite{RS} and \cite{EEL})
and hence present a transformed
Burgers' hierarchy. The following two Propositions give the
result for both the Burgers' hierarchy (\ref{Burg_hier})
and the potential Burgers' hierarchy (\ref{Burg_pot_1}):

\strut\hfill

\noindent
{\bf Proposition 3.2:}
{\it
Under the reciprocal-B\"acklund transformation
\begin{gather}
\label{R_1}
\mbox{$R$}:
\left\{\ba{l}
\displaystyle{dy(x,t)
=\Phi_1^t dx-\Phi_{1,n}^xdt}
\\[3mm]
d\tau(x,t)=dt \\[3mm]
U(y,\tau)=u 
\ea\right.
\end{gather}
with
\begin{gather}
\label{R_1_Phi}
\Phi_1^t=u,\qquad \Phi_{1,n}^x=-P^{n}[u]\,u
\end{gather}
the Burgers' hierarchy
\begin{gather}
\label{RB_Burg}
u_t=R^n[u]\,u_x,\qquad\mbox{where}\ \ R[u]=D_x[u]+u+u_xD_x^{-1}\circ 1,
\end{gather}
transforms to the hierarchy
\begin{gather}
\label{RB_1}
U_\tau=\left\{
\vphantom{\left(e^VD_y[V]\right)^ne^V}
UD_y[U]-U_y\right\}
\left\{
\vphantom{\left(e^VD_y[V]\right)^ne^V}
P^n[U]\,U\right\}\ ,
\end{gather}
where
\begin{gather}
P[U]=UD_y[U]+U.
\end{gather}
}

\noindent
{\bf Proposition 3.3:}
{\it
Under the reciprocal-B\"acklund transformation
\begin{gather}
\label{R_2}
\mbox{$R$}:
\left\{\ba{l}
\displaystyle{dy(x,t)
=\Phi_2^t dx-\Phi_{2,n}^xdt}
\\[3mm]
d\tau(x,t)=dt \\[3mm]
V(y,\tau)=v
\ea\right.
\end{gather}
with
\begin{gather}
\label{R_2_Phi}
\Phi_2^t=e^v,\qquad \Phi_{2,n}^x=-D_x^{n}[v]e^v,
\end{gather}
the potential Burgers' hierarchy
\begin{gather}
\label{RB_Pot_Burg}
v_t=P^n[v_x]\,v_x,\qquad \mbox{where}\ \  P[v_x]=D_x[v_x]+v_x,
\end{gather}
transforms to the hierarchy
\begin{gather}
\label{RB_2}
V_\tau=\left\{
\vphantom{\left(e^VD_y[V]\right)^ne^V}
D_y[V]-V_y\right\}
\left\{\left(
\vphantom{\frac{da}{db}}
e^VD_y[V]\right)^ne^V\right\}.
\end{gather}
}

\strut\hfill

\noindent
We give some explicit examples of the equations (\ref{RB_1})
and (\ref{RB_2}):

\strut\hfill

\noindent
Under the reciprocal-B\"acklund transformation (\ref{R_1})--(\ref{R_1_Phi})
the Burgers' hierarchy  (\ref{RB_Burg})
with $n=1$ transforms to
\begin{gather}
U_\tau=U^2U_{yy}+U^2U_y
\end{gather}
and for $n=2$ we obtain
\begin{gather}
U_\tau=U^3U_{3y}+3U^3U_{yy}
+3U^2U_yU_{yy}+3U^2U_y^2+2U^3U_y.
\end{gather}
Under the reciprocal-B\"acklund transformation
(\ref{R_2})--(\ref{R_2_Phi})
the potential Burgers' hierarchy  (\ref{RB_Pot_Burg})
with $n=1$ transforms to
\begin{gather}
V_\tau=e^{2V}\left(V_{yy}+V^2_y\right)
\end{gather}
and for $n=2$ we obtain
\begin{gather}
V_\tau=e^{3V}\left(V_{3y}+6V_yV_{yy}+4V_y^3\right).
\end{gather}

\section{The Calogero-Degasperis-Ibragimov-Shabat hierarchy}

The third-order evolution equation
\begin{gather}
\label{IS-1}
u_t=u_{3x}+3u^2u_{xx}+9uu_x^2+3u^4u_x
\end{gather}
is known as the Calogero-Degasperis-Ibragimov-Shabat equation and is a
well-known
$C$-integrable evolution equation which can be linearised by a nonlocal
transformation (\cite{Sokolov84}, \cite{Calogero87}, \cite{Petersson}).
In \cite{Petersson} we derived a second-order nonlocal
recursion operator for (\ref{IS-1}), namely
\begin{gather}
R[u]=D_x^2+2u^2D_x+10uu_x+u^4\notag \\
\qquad +2\left(
u_{xx}+2u^2u_x
+2ue^{-2\int u^2\,dx}\int e^{2\int u^2\,dx}\,u_x^2\,dx\right)
D_x^{-1}\circ u\notag\\
\label{recursion_IS}
\qquad -2ue^{-2\int u^2\,dx}D_x^{-1}\circ \left[
\left(u_{xx}+2u^2u_x\right)e^{2\int u^2\,dx}
+2u\int e^{2\int u^2\,dx}\,u_x^2\,dx\right]
\end{gather}
and also reported some nonlocal 
symmetries that follow from this recursion operator.
In terms of the recursion operator (\ref{recursion_IS}) a
local Calogero-Degasperis-Ibragimov-Shabat hierarchy of $C$-integrable 
evolution equations can be presented in the form
\begin{gather}
\label{IS-hier}
u_t=R^n[u]\,u_x, \qquad n=1,2,\ldots\ .
\end{gather}
Equation (\ref{IS-1}) then
corresponds to (\ref{IS-hier}) with
$n=1$. For $n=2$
the second member
of hierarchy (\ref{IS-hier}) is
\begin{gather}
u_t=u_{5x}+5u^2u_{4x}+40uu_xu_{3x}+25uu_{xx}^2+50u_x^2u_{xx}+10u^4u_{3x}
\notag\\[0.3cm]
\label{IS-2nd}
\qquad
+120u^3u_xu_{xx}+140u^2u_x^3
+10u^6u_{xx}+70 u^5u_x^2+5u^8u_x.
\end{gather}


We now investigate the nonlocal symmetry structure in the sense of
its first- and second-degree potential symmetries and obtain the
corresponding nonlocal conservation laws. We show that the
linearisations of
(\ref{IS-1}) and (\ref{IS-2nd}) follow directly from the second
potentialisation of (\ref{IS-1}) and (\ref{IS-2nd}), respectively. 

\subsection{Nonlocal conservation laws and linearisation of the
Calogero-Degas-\n
peris-Ibragimov-Shabat hierarchy}

The results for the first and second members of the hierarchy
(\ref{IS-hier}) are given by the following two propositions:

\strut\hfill

\noindent
{\bf Proposition 4.1:} {\it
The Calogero-Degasperis-Ibragimov-Shabat equation (\ref{IS-1}),
\begin{gather*}
u_t=u_{3x}+3u^2u_{xx}+9uu_x^2+3u^4u_x,
\end{gather*}
admits a first potentialisation of the form
\begin{gather}
\label{IS-Pot-1}
v_t=v_{3x}-\frac{3}{4}\frac{v_{xx}^2}{v_x}+3v_xv_{xx}+v_x^3
\end{gather}
and second potentialisation of the form
\begin{gather}
\label{IS-Pot-2}
w_t=w_{3x},
\end{gather}
where
\begin{subequations}
\begin{gather}
\label{IS1_aux_1}
v_x=u^2\\[0.3cm]
\label{IS1_aux_2}
v_t=2uu_{xx}-u_x^2+6u^3u_x+u^6\\[0.3cm]
\label{IS1_aux_3}
w_x=e^vv_x^{1/2}\\[0.3cm]
\label{IS1_aux_4}
w_t=e^v\left(
\frac{1}{2}v_x^{-1/2}v_{3x}-\frac{1}{4}v_x^{-3/2}v_{xx}^2
+2v_x^{1/2}v_{xx}+v_x^{5/2}\right).
\end{gather}
\end{subequations}
The corresponding
nonlocal conserved current, $\Phi^t$, and flux, $\Phi^x$, are
\begin{subequations}
\begin{gather}
\Phi^t=ue^{\int u^2\,dx}\\
\Phi^x=-\left(
u_{xx}+4u^2u_x+u^5\right)
e^{\int u^2\,dx}
\end{gather}
\end{subequations}
and the linearising transformation that transforms (\ref{IS-1})
to (\ref{IS-Pot-2}) is
\begin{gather}
w_x=ue^{\int u^2\,dx}.
\end{gather}
}

For the second member of the Calogero-Degasperis-Ibragimov-Shabat hierarchy we
have

\strut\hfill

\noindent
{\bf Proposition 4.2:} {\it
The second Calogero-Degasperis-Ibragimov-Shabat equation (\ref{IS-2nd}),
\begin{gather*}
u_t=u_{5x}+5u^2u_{4x}+40uu_xu_{3x}+25uu_{xx}^2+50u_x^2u_{xx}+10u^4u_{3x}
\notag\\[0.3cm]
\qquad
+120u^3u_xu_{xx}+140u^2u_x^3
+10u^6u_{xx}+70 u^5u_x^2+5u^8u_x\ ,
\end{gather*}
admits a first potentialisation of the form
\begin{gather}
v_t=v_{5x}+5v_xv_{4x}
-\frac{5}{2}v_x^{-1}v_{xx}v_{4x}
+10v_x^2v_{3x}
+5v_{xx}v_{3x}+5v_x^{-2}v_{xx}^2v_{3x}\nonumber\\[0.3cm]
\label{IS2-Pot-1}
\qquad
-\frac{5}{4}v_x^{-1}v_{3x}^2
-\frac{35}{16}v_{x}^{-3}v_{xx}^4
-\frac{5}{2}v_{x}^{-1}v_{xx}^3
+\frac{25}{2}v_xv_{xx}^2
+10v_x^3v_{xx}+v_x^5
\end{gather}
and second potentialisation of the form
\begin{gather}
\label{IS2-Pot-2}
w_t=w_{5x},
\end{gather}
where
\begin{subequations}
\begin{gather}
\label{IS2_aux_1}
v_x=u^2\\[0.3cm]
v_t=2uu_{4x}
-2u_xu_{3x}
+u_{xx}^2
+10u^3u_{3x}
+50u^2u_xu_{xx}
+20u^5u_{xx}+70u^4u_x^2\nonumber\\[0.3cm]
\label{IS2_aux_2}
\qquad
+20u^7u_x+u^{10}\\[0.3cm]
\label{IS2_aux_3}
w_x=e^vv_x^{1/2}\\[0.3cm]
w_t=e^v\left(
\frac{1}{2}v_x^{-1/2}v_{5x}-v_x^{-3/2}v_{xx}v_{4x}
+3v_x^{1/2}v_{4x}
-\frac{3}{4}v_x^{-3/2}v_{3x}^2
+\frac{9}{4}v_x^{-5/2}v_{xx}^2v_{3x}\right.\nonumber\\[0.3cm]
\label{IS2_aux_4}
\left.
+2v_{x}^{-1/2}v_{xx}v_{3x}+7v_x^{3/2}v_{3x}
-\frac{15}{16}v_x^{-7/2}v_{xx}^4
+\frac{15}{2}v_x^{1/2}v_{xx}^2
+8v_{x}^{5/2}v_{xx}
+v_x^{9/2}\right).
\end{gather}
\end{subequations}
The corresponding
nonlocal conserved current, $\Phi^t$, and flux, $\Phi^x$, are
\begin{subequations}
\begin{gather}
\Phi^t=ue^{\int u^2\,dx}\\
\Phi^x=-\left(
u_{4x}+26uu_xu_{xx}+6u^2u_{3x}
+8u_x^3+44u^3u_x^2\right.\nonumber\\[0.3cm]
\qquad\left.
+14u^4u_{xx}
+16u^6u_x+u^9
\right)\,
e^{\int u^2\,dx}
\end{gather}
\end{subequations}
and the linearising transformation that transforms (\ref{IS-2nd})
to (\ref{IS2-Pot-2}) is
\begin{gather}
w_x=ue^{\int u^2\,dx}.
\end{gather}
}

\strut\hfill

In order to derive the auxiliary systems for the
Calogero-Degasperis-Ibragimov-Shabat hierarchy, (\ref{IS-hier}),
we need the integrating factors of this hierarchy
and the integrating factors of the corresponding potential
hierarchy. For the first two members of the hierarchy the integrating
factors are given by the following

\strut\hfill

\noindent
{\bf Lemma 4.1:} {\it The third-order
Calogero-Degasperis-Ibragimov-Shabat equation, (\ref{IS-1}),
and the fifth-order Calogero-Degasperis-Ibragimov-Shabat equation,
(\ref{IS-2nd}),
admit only one integrating factor, $\Lambda$, namely
\begin{gather}
\Lambda(x,u,u_x,\ldots)=u.
\end{gather}
For the first potentialisation of (\ref{IS-1}), namely for the third-order
potential equation (\ref{IS-Pot-1}),
the complete set of 
integrating factors of second-order are
\begin{gather}
\Lambda(x,v,v_x,v_{xx})=a(x)e^vv_x^{-3/2}v_{xx}
+2a(x)e^vv_x^{1/2}-2e^vv_x^{-1/2}\frac{da}{dx},
\end{gather}
where
\begin{gather}
\frac{d^3a}{dx^3}=0\ ,
\end{gather}
and for the first potentialisation of (\ref{IS-2nd}), namely for the
fifth-order potential equation (\ref{IS2-Pot-1}), the complete
set of integrating factors of second-order are
\begin{gather}
\Lambda(x,v,v_x,v_{xx})=a(x)e^vv_x^{-3/2}v_{xx}
+2a(x)e^vv_x^{1/2}-2e^vv_x^{-1/2}\frac{da}{dx},
\end{gather}
where
\begin{gather}
\frac{d^5a}{dx^5}=0.
\end{gather}
}
To prove Lemma 4.1 we just verify (\ref{Int_lamb}).

\strut\hfill

Lemma 4.1 tempts us to make the following

\strut\hfill

\noindent
{\bf Supposition 4.1:} {\it
All first potentialisations of the
Calogero-Degasperis-Ibragimov-Shabat hierarchy, (\ref{IS-hier}),
for all $n\in {\cal N}$ admit the following complete set
of integrating factors of second order:
\begin{gather}
\label{Lambda_pot_n}
\Lambda(x,u,u_x,u_{xx};\, n)
=a(x)e^vv_x^{-3/2}v_{xx}
+2a(x)e^vv_x^{1/2}-2e^vv_x^{-1/2}\frac{da}{dx},
\end{gather}
where
\begin{gather}
\frac{d^na}{dx^n}=0.
\end{gather}
}


\noindent
{Remark on the proof of Proposition 4.1 and Proposition 4.2:}\n
For the linearisations of (\ref{IS-1}) and (\ref{IS-2nd})
in (\ref{IS-Pot-2}) and (\ref{IS2-Pot-2}),
respectively, we make use of the
integrating factor
\begin{gather}
\Lambda=e^{v}v_x^{-3/4}v_{xx}+2e^{v}v_x^{1/2},
\end{gather}
which corresponds to the case $a(x)=1$
in Lemma 4.1.
If one uses instead the explicitly $x$-dependent integrating
factors, the resulting linear equations also depend
explicitly on $x$.

\subsection{Potential symmetries of the Calogero-Degasperis-Ibragimov-Shabat hierarchy}

We now study the symmetry properties of the auxiliary systems for the
Calogero-Degasperis-Ibragimov-Shabat hierarchy (\ref{IS-hier}). 

\strut\hfill

\noindent
{\bf Case $n=1$:} The first-degree
potential symmetries of the first member of the
hierarchy (\ref{IS-hier}), i.e. (\ref{IS-1}), are given by the Lie point symmetries of the
first auxiliary system (\ref{IS1_aux_1})--(\ref{IS1_aux_2}). We obtain
\begin{subequations}
\begin{gather}
Z_1=\pde{\ }{t},\qquad Z_2=\pde{\ }{x},\qquad Z_3=\pde{\ }{v}\\[0.3cm]
Z_4=\frac{1}{3}x\pde{\ }{x}+t\pde{\ }{t}
-\frac{1}{6}u\pde{\ }{u},\qquad
Z_5=ue^{-2v}\pde{\ }{u}-e^{-2v}\pde{\ }{v}.
\end{gather}
\end{subequations}

\noindent
The second-degree potential symmetries of the first
member of the
hierarchy (\ref{IS-hier}) are given by the Lie point symmetries of the
combined auxiliary system (\ref{IS1_aux_1})--(\ref{IS1_aux_4}). We obtain
\begin{subequations}
\begin{gather}
Z_1=\pde{\ }{t},\qquad Z_2=\pde{\ }{x},\qquad Z_3=\pde{\ }{w}\\[0.3cm]
Z_4=\frac{1}{3}x\pde{\ }{x}+t\pde{\ }{t}
-\frac{1}{6}u\pde{\ }{u}+\frac{1}{6}w\pde{\ }{w},\qquad
Z_5=\pde{\ }{v}+w\pde{\ }{w}\\[0.3cm]
Z_6=\left(\frac{1}{2}e^{-v}-uwe^{-2v}\right)\pde{ }{u}
+we^{-2v}\pde{ }{v}
+\frac{1}{2}x\pde{ }{w}\\[0.3cm]
Z_7=ue^{-2v}\pde{\ }{u}-e^{-2v}\pde{\ }{v}.
\end{gather}
\end{subequations}

\noindent
{\bf Case $n=2$:} The first-degree
potential symmetries of the second member of the
hierarchy (\ref{IS-hier}), i.e. (\ref{IS-2nd}), are given by the
Lie point symmetries of the first auxiliary system
(\ref{IS2_aux_1})--(\ref{IS2_aux_2}). We obtain
\begin{subequations}
\begin{gather}
Z_1=\pde{\ }{t},\qquad Z_2=\pde{\ }{x},\qquad Z_3=\pde{\ }{v}\\[0.3cm]
Z_4=\frac{1}{5}x\pde{\ }{x}+t\pde{\ }{t}
-\frac{1}{10}u\pde{\ }{u},\qquad
Z_5=ue^{-2v}\pde{\ }{u}-e^{-2v}\pde{\ }{v}.
\end{gather}
\end{subequations}
The complete set of Lie point symmetries of
the auxiliary system (\ref{IS2_aux_1})--(\ref{IS2_aux_4}) are
\begin{subequations}
\begin{gather}
Z_1=\pde{\ }{t},\qquad Z_2=\pde{\ }{x},\qquad Z_3=\pde{\ }{w}\\[0.3cm]
Z_4=\frac{1}{5}x\pde{\ }{x}+t\pde{\ }{t}
-\frac{1}{10}u\pde{\ }{u}+\frac{1}{10}w\pde{\ }{w},\qquad
Z_5=\pde{\ }{v}+w\pde{\ }{w}\\[0.3cm]
Z_6=ue^{-2v}\pde{\ }{u}-e^{-2v}\pde{\ }{v}.
\end{gather}
\end{subequations}
We  note that the second member of the hierarchy (\ref{IS-2nd})
does not admit second-degree potential symmetries.

\strut\hfill

We allow ourselves the following

\strut\hfill

\noindent
{\bf Supposition 4.2:} {\it
There exist no second-degree potential symmetries for the
Calogero-Degasperis-Ibragimov-Shabat hierarchy,
(\ref{IS-hier}), for $n>1$ and
the maximum set of first-degree potential symmetries for 
the hierarchy (\ref{IS-hier})
is given by the following Lie symmetry generators:
\begin{subequations}
\begin{gather}
Z_1=\pde{\ }{t},\qquad Z_2=\pde{\ }{x},\qquad Z_3=\pde{\ }{v}\\[0.3cm]
Z_4=\frac{1}{n}x\pde{\ }{x}+t\pde{\ }{t}
-\frac{1}{2n}u\pde{\ }{u},\qquad
Z_5=ue^{-2v}\pde{\ }{u}-e^{-2v}\pde{\ }{v},
\end{gather}
\end{subequations}
for all natural numbers $n>1$.
}

\section{Concluding remarks}
We have introduced second-degree potential symmetries in Definition
2.1 and studied the Burgers' hierarchy and the
Calogero-Degasperis-Ibragimov-Shabat hierarchy.
We obtained second-degree potential symmetries only for
the first members of the hierarchies.
Nonlocal conservation
laws and nonlocal transformations which linearise the hierarchies
were obtained through the second potentialisations.
It would be interesting to investigate higher-degree
potential symmetries further for other symmetry-integrable hierarchies
and linearisable hierarchies as well as for systems of evolution
equations. Preliminary calculations show that systematic
potentialisation of some symmetry-integrable equations, such as the
Krichever-Novikov equation, lead to interesting
auto-B\"acklund transformations for the equations. A complete description
of the potentialisation for a class of Krichiver-Novikov equations and
its connection with higher-degree potential symmetries is currently
in preparation \cite{NM2} and is planned for publication as a
follow-up paper.

\section*{Acknowledgement}
ME acknowledges the financial support provided by the LTU
grant nr. 2557-05.

\end{document}